\documentclass{article}

\usepackage{PRIMEarxiv}
\usepackage[utf8]{inputenc} 
\usepackage[T1]{fontenc}    
\usepackage{hyperref}       
\usepackage{url}            
\usepackage{booktabs}       
\usepackage{amsfonts}       
\usepackage{nicefrac}       
\usepackage{microtype}      
\usepackage{lipsum}
\usepackage{amsmath,amssymb}
\usepackage{chngcntr}
\usepackage{graphicx}
\graphicspath{{media/}}     

\title{Detect influential points of feature rankings}

\author{
  Shuo Wang \\
  Institute of Medical Biometry and Statistics \\
  Faculty of Medicine and Medical Center \\
  University of Freiburg \\
  Germany\\
  \texttt{shuo.wang@uniklinik-freiburg.de} \\
   \And
  Junyan Lu 
  \thanks{Corresponding author: Junyan Lu, junyan.lu@uni-heidelberg.de} \\
  Medical Faculty Heidelberg \\
  Heidelberg University \\
  Germany\\
  \texttt{junyan.lu@uni-heidelberg.de} 
}

\begin{document}
\maketitle

\begin{abstract}
\textbf{Background}  
Deriving feature rankings is essential in bioinformatics studies since the ordered features are important in guiding subsequent research. Feature rankings may be distorted by influential points (IP), but such effects are rarely mentioned in previous studies. This study aimed to investigate the impact of IPs on feature rankings and propose a new method to detect IPs. 
\textbf{Method} 
The present study utilized a case-deletion (i.e., leave-one-out) approach to assess the impact of cases. The influence of a case was measured by comparing the rank changes before and after the deletion of that case. We proposed a rank comparison method using adaptive top-prioritized weights that highlighted the rank changes of the top-ranked features. The weights were adjustable to the distribution of rank changes.\\
\textbf{Results} 
Potential IPs could be observed in several datasets. The presence of IPs could significantly alter the results of the following analysis (e.g., enriched pathways), suggesting the necessity of IPs detection when deriving feature rankings. Compared with existing methods, the novel rank comparison method could identify rank changes of important (top-ranked) features because of employing the adaptive weights adjusted to the distribution of rank changes. 
\textbf{Conclusions} 
IPs detection should be routinely performed when deriving feature rankings. The new method for IPs detection exhibited favorable features compared with existing methods.  

\end{abstract}


\section{Introduction}

The high-throughput sequencing enables us to measure numerous features simultaneously. However, the vast majority of the features are irrelevant to the interest of research \cite{Churko.2013}. Generally, researchers will concentrate on a small subset of important features that are associated with research objectives \cite{WietlickaPiszcz.2013}. To identify these important features, we can order these features according to their marginal correlation with outcomes and extract the important ones based on a predefined criterion, such as a predefined significant level \cite{Boulesteix.2009, Li.2011, Shi.2007, Aerts.2006, Pham.2006}.

The obtained feature ranking is important in guiding the following research \cite{WietlickaPiszcz.2013}. It provides valuable insights into the underlying biological processes and molecular mechanisms involved \cite{Heydebreck.2004, Boulesteix.2009}. Additionally, the feature list itself can be utilized as input for further analyses. For instance, Gene Set Enrichment Analysis (GSEA) is a powerful and classical analytical method to identify functional pathways that are associated with the disease phenotype \cite{Subramanian.2005, Shi.2007}. The enrichment scores of the pathways are calculated by walking down the ordered feature lists, increasing a running sum statistic if the gene is in the tested gene set, and decreasing it otherwise. 
These analyses are based on the assumption that the derived feature rankings are stable, which is yet typically impractical, particularly in the context of bioinformatics where sample sizes are typically small \cite{Nurunnabi.2014, WietlickaPiszcz.2013, He.2010}. Ranking stability refers to the degree of insensitivity of the rankings to data perturbation, such as the deletion of a case \cite{Alelyani.}. Ideally, we expect a ranking remains approximately the same since the removal of a few cases. However, a feature list may differ substantially due to data perturbation \cite{Boulesteix.2009, Zyla.2017}.

Generally, there are two approaches to enhancing the stability of feature lists. The first strategy is to aggregate feature lists obtained with resampled samples \cite{Boulesteix.2009, Gawe.2013, He.2010, Li.2019}. Boulesteix et.al. stated that “aggregation, in particular, aggregation of results obtained from slightly perturbed versions of the data set, is expected to yield more accurate rankings”\cite{Boulesteix.2009}.  The challenge of this approach is that the aggregated features can be affected by the size of the resampling \cite{Alelyani.}. A small resampling size, say selecting 60\% of the cases, can yield extra instability since the resampled data significantly differs from the original data; however, a large resampling size, say 90\% of cases, means a poor improvement in stability because the extreme cases are included in 90\% of the subsets. Therefore, the aggregation strategy can enhance stability but fails to handle the instability caused by extreme cases.

Another solution to strengthen the stability is to mitigate the effects of extreme values for each feature. For instance, we can truncate the extreme values of each feature \cite{Anders.2012}. This approach assumes that all values far away from the rest are extreme values, which is definitely incorrect \cite{Zimek.2012}. Instead of handling the extreme values for each feature, a more pragmatic solution is to check for the cases that heavily influence the overall feature rankings. As such, the stability of feature ranking is enhanced with the minimum alteration of the data structure. These cases with a disproportionate influence than the rest are termed influential points (IPs) \cite{Everitt.2002}.

IPs detection should be routinely conducted when deriving feature rankings. However, such a process is still rarely mentioned because of the insufficient awareness of the detrimental effects of IPs and a lack of computational tools and guidelines to detect them. Therefore, the primary objective of the study is to present a simple approach for detecting IPs of feature rankings. Our strategy is based on case-deletion, which is used by many IPs detection methods for regression modeling, such as Cook’s distance and DFFITs \cite{Cook.1977, Belsley.2005}. The effects of a single case on feature rankings are quantified by measuring the weighted rank changes that the case incurred. We also proposed a rank-comparing measure that utilizes adaptive weights to highlight the rank changes of the top-ranked features. The secondary objective is to demonstrate the potential impact of IPs in real-world data.

In the methods section, \textbf{\ref{sec:2.1}} provided information about the used datasets and the preprocessing procedure. \textbf{\ref{sec:2.2}} and \textbf{\ref{sec:2.3}} demonstrated the approaches to deriving feature rankings and a brief summary for the procedure of IPs detection. In \textbf{\ref{sec:2.4}}, we illustrated a rank comparison method using adaptive weights. \textbf{\ref{sec:2.5}} provided a package for the new method and codes for result reproducibility. \textbf{\ref{sec:3.1}} in the results section presented an example to illustrate the IPs detection procedure using the new method, and \textbf{\ref{sec:3.2}} showed the relationship between the choice of weights and the distribution of rank changes. \textbf{\ref{sec:3.3}} exhibited the results of IPs detection for all datasets. \textbf{\ref{sec:3.4}} compared the results with the other two existing rank comparison methods. \textbf{\ref{sec:3.5}} exhibited the effects of IPs on the subsequent analyses.

\section{Methods}

\subsection{Datasets}
\label{sec:2.1}

All analyses were performed on ten public-accessible mRNA expression profiles derived from The Cancer Genome Atlas (TCGA, \url{https://www.cancer.gov/tcga}). TCGA enrolled more than 11,000 cases across 33 tumor types, providing various omics information, including mRNA, single nucleotide polymorphisms, methylation, etc. The sufficient sample size and rich information make it a popular database that has been widely analyzed. A large number of publications have been generated by analyzing the TCGA datasets; however, IPs were rarely investigated \cite{Hutter.2018}. Feature rankings were derived by comparing the gene expression of the primary solid tumor tissues (coded as 01 in TCGA) and the solid normal tissues (coded as 11). A challenge of TCGA data is that it collected much more tumor samples than the normal control samples. For this sake, we removed several datasets due to insufficient normal controls and finally included ten datasets. For the included datasets, we randomly drew tumor cases with twice the number of the normal cases forming the new tumor group in order to balance the sample sizes between the two groups.

We normalized the expression counts into counts per million (CPM) and scaled by trimmed mean of M values (TMM) for these included mRNA expression profiles \cite{Chen.2014}. We removed i) the low-expressed features whose expressions were zero in more than half of the samples; ii) and the features coded by Ensembl IDs without corresponding Gene Symbols. The information for the considered datasets is summarized in Table \ref{tab:1}. 

\begin{table}
\caption{summary of the data}
\centering
\begin{tabular}{lllll}
\toprule
Study & Description                           & Cancer & Normal & Features \\
\midrule
BRCA  & Breast invasive carcinoma             & 226    & 113    & 14543    \\
COAD  & Colon adenocarcinoma                  & 82     & 41     & 14181    \\
HNSC  & Head and Neck squamous cell carcinoma & 88     & 44     & 13805    \\
KIRC  & Kidney renal clear cell carcinoma     & 144    & 72     & 14549    \\
LIHC  & Liver hepatocellular carcinoma        & 100    & 50     & 13095    \\
LUAD  & Lung adenocarcinoma                   & 118    & 59     & 14645    \\
LUSC  & Lung squamous cell carcinoma          & 98     & 49     & 14645    \\
PRAD  & Prostate adenocarcinoma               & 104    & 52     & 14584    \\
STAD  & Stomach adenocarcinoma                & 64     & 32     & 14595    \\
THCA  & Thyroid carcinoma                     & 116    & 58     & 14374    \\
\bottomrule
\end{tabular}
\label{tab:1}
\end{table}

\subsection{Feature rankings}
\label{sec:2.2}

For binary response data, fold change and t-test are two popular strategies to extract important features due to their simplicity. For fold change, we simply order the features according to the ratio of means between two groups, with the larger (or smaller) ratio indicating a closer connection to the outcomes. For the t-test, we ascendingly order the features according to the significance of the t-test. The important features are those with smaller P values. The two strategies are straightforward, but both have weaknesses. Fold change is subject to bias since it does not consider the feature variance. t-test cannot accurately estimate the variance without sufficient samples \cite{Cui.2003}. \textit{edgeR} and \textit{DESeq} are popular packages for gene differential expression; both of them are grounded on a negative binomial distribution assumption and use the exact test analogous to Fisher’s exact test to estimate the difference between groups \cite{Robinson.2010, Anders.2010}. \textit{voom} and \textit{DESeq} are two popular packages based on Bayes \cite{Law.2014, Leng.2013}. Since the methods for feature ranking is not the primary interest, the present study used the t-test to derive feature rankings for its simplicity. 

\subsection{The IPs detection method for feature rankings}
\label{sec:2.3}

The IPs detection method is proposed based on case deletion (e.g., leave-one-out), which is also the foundation of many classical methods, such as COOK’s distance and DFFITS \cite{Cook.1977, Belsley.2005}. The idea is to compare the results of original and leave-one-out data. The influence of a case is quantified by the total rank changes caused by the deletion of that case \cite{Rajaratnam.2019}. Potential IPs are the ones inducing aberrant rank changes. The procedure can be detailed in three steps:

\textbf{Step 1}, generate the original ranking and leave-one-out rankings using a feature ranking method, such as t-test. For the data with n case, we can obtain one original and n leave-one-out feature rankings. \\
\textbf{Step 2}, calculate rank changes. It is recommended to use top-prioritized weights in the rank comparison (see \ref{sec:2.4.1} and \ref{sec:2.4.2} \\
\textbf{Step 3}, calculate the total rank changes for each observation(see \ref{sec:2.4.3}). Perform diagnostic checks to identify potential influential points (and \ref{sec:3.3}). 

\subsection{Rank comparison methods}
\label{sec:2.4}

The influence of a case is quantified by calculating the rank changes with a suitable rank comparison method. Spearman and Kendall rank correlation coefficient are two well-known methods for rank comparison; however, both assume that all elements in the rank lists are of equal importance, which may not be appropriate in real-world applications \cite{Abdi.2007, Sedgwick.2014}. Naturally, the top-ranked features are more critical \cite{Li.2011}: firstly, the top-ranked elements are more relevant to the outcomes \cite{WietlickaPiszcz.2013}; secondly, features in the top positions are more stable. Therefore, it is important to prioritize the top-ranked features by allocating more weights to them.

The challenge lies in selecting appropriate weights. Several weight methods have been proposed based on the idea of weighting more to the top. For instance, Stillwell et al. proposed rank reciprocal (RR) weights $(\frac{1}{r})$ and rank sum (RS) weights \cite{Stillwell.1981}, Solymosi et al. proposed rank order centroid (ROC) weights $\frac{100\sum_{i=r}^{n}\frac{1}{i}}{\sum_i^n\frac{1}{i}}$, where $n$ and $r$ denote the number of features and the ranks \cite{Solymosi.1986}. Although, these weight methods are top-prioritized, but not adjustable for different data, i.e., weights are fixed. It is unreasonable to allocate an unstable rank list with the same type of weights as a stable one. A preferable weight method should be adaptive for different data.

In addition, it is advisable to have the weights standardized for better interpretability of results. For instance, if the weight for a specific rank is 0.01, we can easily understand that this rank takes up 1\% of the total weight, which can be an indicator of the feature's importance.

Here, we proposed a rank comparison method that uses standardized, monotonically decreasing and data-adaptive weights.

\subsubsection{Weight function}
\label{sec:2.4.1}

The present weight method is based on the reciprocal exponential function ($w_x=\lambda e^{-\kappa x}$). As such, weights are positive and decreasing \cite{Yang.2006}. We extend the exponential weights by i) incorporating the length of lists into the weight function; ii) limiting the total weights to one. After simplification, the weight function is given by,

\begin{equation}
    w_x = - \frac{1-e^\kappa}{1-e^{-\kappa m}}e^{-\kappa x}
    \label{eq:1}
\end{equation}

where $x$ is the ranks, m is the length of feature lists, and $\kappa$ determines the shape of the weight curve. We can deduce from the formula that i) the total weights always converge to 1 for any given $m$, e.g., 1 or $+\infty$, or $\kappa$. In other words, weights are standardized; ii) a large m indicates smaller weights for each rank; iii) a large $\kappa$ means weighting more to the top. 

\begin{figure}
    \centering
    \includegraphics[width=0.65\textwidth]{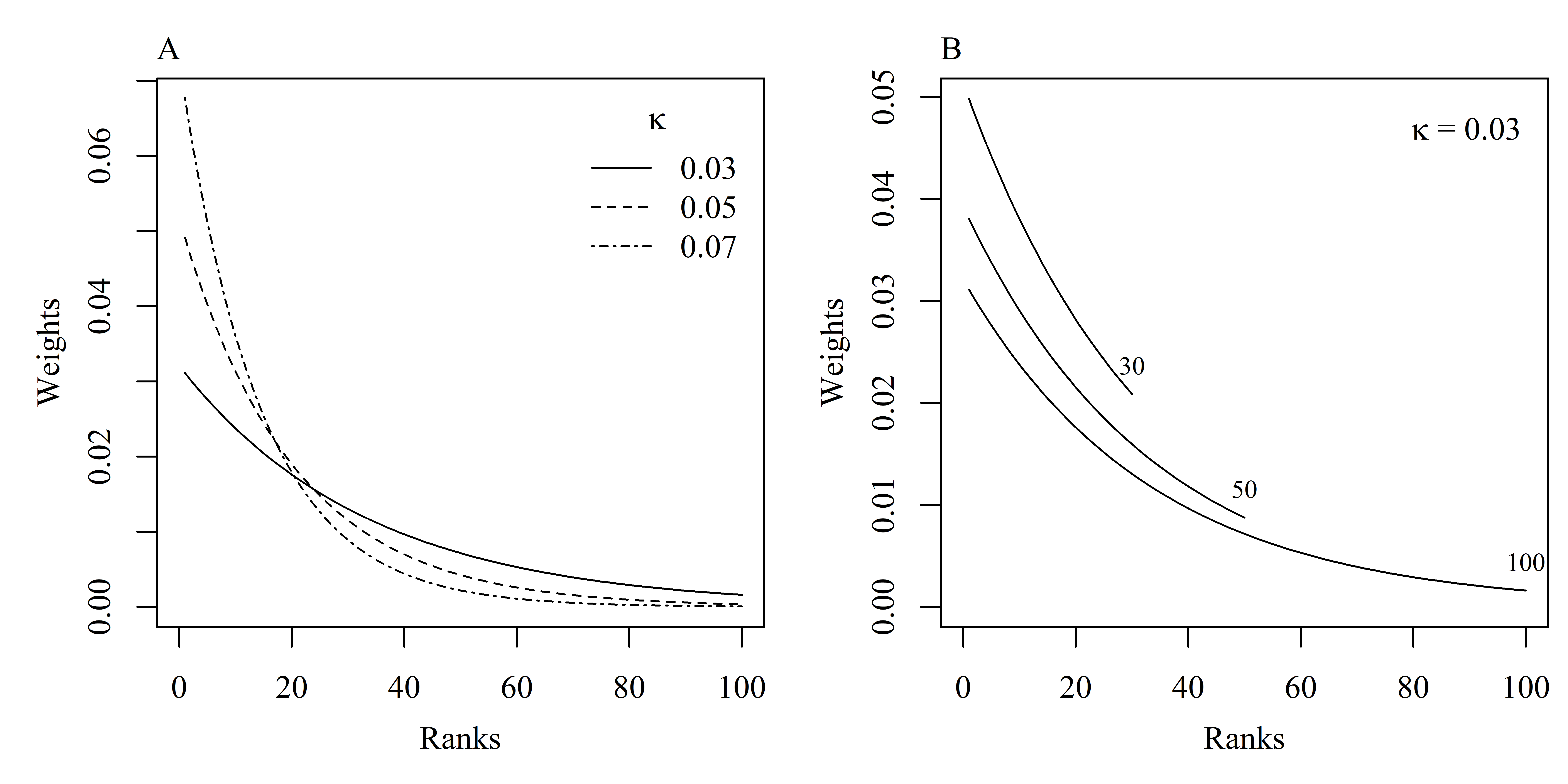}
    \caption{the behavior of the weight function using different $\kappa$ (Panel A) and lengths of the list (Panel B).}
    \label{fig:1}
\end{figure}

Figure \ref{fig:1} depicts the role of $kappa$ and $m$ in the weight function. Note that weights in Figure 1 should be discretized points, but we plotted them as a continuous curve in order to show the decreasing trend. As is shown, a large $\kappa$ corresponds to a steeper curve and more weight to the top. Since the total weight is one regardless of the $\kappa$, more weights to the top means fewer weights to the tail. Weights also depend on the length of the ranking lists. Features of a long list share fewer weights than a short list. For the two parameters in the function, m is typically given, and $\kappa$ is the only unknown parameter. Subjective knowledge should be the highest priority in determining $\kappa$; however, they are not always available. To address this issue, we presented an approach that $\kappa$ was selected in a data-driven way.

\subsubsection{Choice of $\kappa$}
\label{sec:2.4.2}

Assume a data matrix $(X_{n,m})$ with n observations and m features. $r_{0,j \in 1:m}$ and $r_{i \in 1:n,j \in 1:m}$ are original ranks and leave-one-out ranks. With (\ref{eq:1}), the ranks are transformed to their corresponding weights. Hereinafter, we call them weighted ranks. The influence of a case is quantified by the weighted ranks changes between $r_0$ and $r_{i \in 1:n}$. A critical issue is choosing a proper $\kappa$ in (\ref{eq:1}). An arbitrary $\kappa$ may exaggerate the rank changes for some ranks while underrating the rank changes for others. Typically, a parameter can be optimized by maximizing or minimizing a particular statistic. Here, we proposed that $\kappa$ can be optimized by maximizing the similarity between the weighted original and leave-one-out ranks. More precisely, we select all changed ranks in leave-one-out rankings $(r_{i,j} \mid r_{0,j} \neq r_{i,j})$ and their corresponding original ranks $(r_{0,j} \mid r_{0,j} \neq r_{i,j})$, forming two lists $x$ and $x'$, with the latter being the original list. In essence, the process of maximizing the similarity between the two lists is to adjust the weights to the optimal in which the effects of rank changes are minimized. To this end, we need to transform $x$ and $x'$ into weighted ranks using (\ref{eq:1}), yielding $f(w \mid \kappa,x)$ and $f(w'\mid \kappa,x')$ with an unknown parameter $\kappa$ inside.

\begin{equation}
f(w\mid\kappa, x) = - \frac{1 - e^\kappa}{1 - e^{-\kappa m}} e^{-\kappa x}
\label{eq:2}
\end{equation}

\begin{equation}
f(w'\mid \kappa, x') = - \frac{1 - e^\kappa}{1 - e^{-\kappa m}} e^{-\kappa x'}
\label{eq:3}
\end{equation}

The relation between $f(w'\mid \kappa, x')$ and $f(w \mid \kappa,x)$ can be expressed as

\begin{equation}
 f(w \mid \kappa,x) = f(w'\mid \kappa, x') + \epsilon
\label{eq:4}
\end{equation}

where $\epsilon$ is the weighted rank changes. To derive a proper $\kappa$, we maximize the similarity of the weighted rank changes between $ f(w \mid \kappa,x)$ and $ f(w'\mid \kappa, x') $. (\ref{eq:4}) is alike a model with $f(w \mid \kappa,x)$ being the observed response and $f(w'\mid \kappa,x')$ being the predicted values. The maximum similarity can be achieved by maximizing the $R^2$ assuming $f(w'\mid \kappa,x')$ and $f(w \mid \kappa,x)$ are the predicted and observed responses. 

\begin{equation}
   \mathop{\arg \max}\limits_{\kappa} = 1 - \frac{(f(w' \mid \kappa, x') - f(w \mid \kappa, x))^2}{(f(w \mid \kappa, x) - \overline{ f(w \mid \kappa, x)})^2}
   \label{eq:5}
\end{equation}

$(f(w' \mid \kappa, x') - f(w \mid \kappa, x))^2$ is the square of the weighted rank changes ($\epsilon$), and $(f(w \mid \kappa, x) - \overline {f(w \mid \kappa, x)})^2$ represents the discriminability of the ranks. (\ref{eq:5}), in principle, minimizes the rank changes while maintaining the discriminability between the ranks. Since $\kappa$ is optimized using all rank changes of n leave-one-out rankings, it can be considered an average decision; therefore, the resulting $\kappa$ is fair for each list. 

Of note, minimizing the rank changes ($\epsilon$) without considering the discriminability can result in an unreasonable $\kappa$. In fact, the minimized $\epsilon$ is always observed at $\kappa = 0$. According to L’ Hospital’s rule, $ \mathop{\lim} \limits_{\kappa \rightarrow 0} f(w \mid \kappa, x) = \frac{1}{m}$, in other words, $\kappa = 0$ corresponds to equal weights. The weighted rank changes $|\epsilon| \rightarrow 0$ under equal weights. It explains why $\kappa$ is optimized at 0 if we minimize $\epsilon$. Equal weights conflict with our preference to weight more to the top; thus, $\kappa$ cannot be optimized by directly minimizing rank changes; instead, the optimal solution should minimize rank changes while maintaining fair discriminability between ranks.

\subsubsection{Total rank changes}
\label{sec:2.4.3}

The total rank changes are defined as the sum of weighted rank changes of all features. The rank change of a feature is usually computed with the rank changes in the absolute form. Here, we used the squared form in order to align with the weight function, which was also optimized in the squared form. The total weighted rank change for the \textit{i-th} case deletion is given by

\begin{equation}
    R_i = \sum_{j = 1}^m(f(w'_j \mid \kappa, r_{0, j}) - f(w_j \mid \kappa, r_{i,j}))^2
\label{eq:6}
\end{equation}

where $m$ is the number of features, $r_{0,j}$ and $r_{i,j}$ are the ranks of the \textit{j-th} features in the original ranking and \textit{i-th} leave-one-out ranking. The case generating the maximum $R_i$ is suspected to be the potential IP. It is worth noting that this method can only check for one observation because of the use of leave-one-out, in which the results merely reflect the effects of a single case. For better presentation, we standardized $R_i$ when they were presented in figures. 

\subsubsection{Compare with the other methods}
\label{sec:2.4.4}

The existing methods proposed for rank comparison can be classified into two categories according to whether weights are considered. The Spearman’s coefficient is a well-known unweighted rank comparison method whose formula is $r_s = 1 - \frac{6\sum_{i = 1}^{m}(R_i - Q_i)^2}{m^3 - m}$, where R and Q are two rank lists containing m elements. Costa et al. extended Spearman’s coefficient by incorporating weights \cite{DaPintoCosta.2005} . The weighted version of correlation is given by $r_w = 1 - \frac{6\sum_{i = 1}^{m}(R_i - Q_i)^2(m - R_i + 1) + (m - Q_i + 1)}{m^4 + m^3 - m^2 - m}$, where the term $(R_i - Q_i)^2$ is exactly Spearman’s method, and it is weighted by the following terms containing the position of the two ranks ($(m-R_i + 1)+ (m - Q_i + 1)$). One problem is that the weights are only related to the two ranks, and it does not incorporate the specific information of the data. In other words, given $R_i$, $Q_i$, and m, the weights are the same regardless of the data.

We compared our method to the classical Spearman’s coefficient and the weighted extension. To simplify the calculation, we can compare the $D_s = \sum_{i = 1}^{m}(R_i - Q_i)^2$  for Spearman’s coefficient and $D_w = \sum_{i = 1}^{m}(R_i - Q_i)^2((m - R_i + 1) + (m - Q_i + 1)$ for the weighted version, since all rankings are equal in length, hence denominators can be removed. We standardized the results by dividing the standard deviation in order to put them on the same scale.

\subsection{Software and reproducibility}
\label{sec:2.5}

The present method for IPs detection of feature rankings was implemented in an R package, which was accessible from (\url{https://github.com/ShuoStat/IPSR}). The data used in this study and codes for results reproducing were available at (\url{https://github.com/ShuoStat/IPs_on_feature_rankings}).

\section{Results}
\subsection{IPs detection procedure}
\label{sec:3.1}
We used LUSC data as an example to illustrate the procedure of IPs detection for feature rankings.

\subsubsection{Derive original ranking and leave-one-out rankings (Step1)}
\label{sec:3.1.1}
The first step of the IPs detection is to derive the original rankings and leave-one-out rankings. Genes were ordered ascendingly according to the p-values of the t-test. We included the top 200 important genes from the original data since the primary interests commonly lie in the top-ranked genes. 

Table \ref{tab:2} shows an example concerning the original and leave-one-out rankings. The LUSC data contained a total of 147 observations, resulting in 147 leave-one-out rankings. From the first ten observations exhibited in Table \ref{tab:2}, we could observe that the top-ranked features were more stable to data perturbation (i.e., case deletion). The orders of the top three genes (CDCA8, AUNIP, and KIF2C) remained unchanged for the first ten observations. More rank changes could be found with the increase of ranks. This is one of the reasons that we weight more to the top-ranked features. 

\begin{table}[]
\centering
\caption{an example to show the original ranking and leave-one-out rankings using LUSC data. The table shows the ranks of the top 20 features and the first ten leave-one-out rankings. }
\label{tab:2}
\begin{tabular}{llllllllllll}
\toprule
Genes   & Original & obs 1 & obs 2 & obs 3 & obs 4 & obs 5 & obs 6 & obs 7 & obs 8 & obs 9 & obs 10 \\
\midrule
CDCA8   & 1        & 1     & 1     & 1     & 1     & 1     & 1     & 1     & 1     & 1     & 1      \\
AUNIP   & 2        & 2     & 2     & 2     & 2     & 2     & 2     & 2     & 2     & 2     & 2      \\
KIF2C   & 3        & 3     & 3     & 3     & 3     & 3     & 3     & 3     & 3     & 3     & 3      \\
CDCA5   & 4        & 4     & 5     & 4     & 4     & 4     & 4     & 4     & 4     & 4     & 4      \\
CENPA   & 5        & 5     & 4     & 5     & 5     & 5     & 5     & 5     & 5     & 5     & 5      \\
TPX2    & 6        & 6     & 6     & 6     & 6     & 6     & 6     & 6     & 6     & 6     & 6      \\
TROAP   & 7        & 7     & 9     & 7     & 7     & 7     & 7     & 7     & 7     & 7     & 7      \\
TFAP2A  & 8        & 9     & 11    & 9     & 9     & 9     & 8     & 8     & 8     & 8     & 10     \\
PLK1    & 9        & 8     & 12    & 11    & 8     & 8     & 9     & 9     & 9     & 9     & 9      \\
BIRC5   & 10       & 11    & 10    & 14    & 11    & 10    & 12    & 10    & 11    & 10    & 8      \\
NUF2    & 11       & 12    & 8     & 8     & 12    & 11    & 13    & 11    & 10    & 11    & 14     \\
UBE2C   & 12       & 10    & 16    & 12    & 14    & 14    & 14    & 12    & 13    & 15    & 15     \\
CCNB1   & 13       & 14    & 7     & 13    & 16    & 15    & 15    & 13    & 12    & 13    & 11     \\
RAMP2   & 14       & 15    & 21    & 17    & 13    & 13    & 10    & 15    & 16    & 12    & 13     \\
TTK     & 15       & 16    & 15    & 10    & 15    & 12    & 11    & 14    & 15    & 16    & 16     \\
NOSTRIN & 16       & 17    & 19    & 15    & 10    & 16    & 16    & 16    & 17    & 14    & 17     \\
KIF4A   & 17       & 18    & 17    & 19    & 17    & 18    & 17    & 17    & 14    & 17    & 12     \\
HJURP   & 18       & 19    & 20    & 18    & 19    & 20    & 18    & 18    & 18    & 18    & 18     \\
ACVRL1  & 19       & 20    & 18    & 16    & 18    & 19    & 19    & 19    & 20    & 20    & 19     \\
PRC1    & 20       & 13    & 13    & 21    & 20    & 17    & 20    & 20    & 19    & 19    & 20     \\
\vdots       & \vdots         & \vdots      & \vdots     & \vdots     & \vdots     & \vdots      & \vdots     & \vdots     & \vdots    &\vdots    & \vdots  \\
\bottomrule
\end{tabular}
\end{table}

\subsubsection{Transform ranks to adaptive weighted ranks (step 2)}

In the second step, the original ranks were converted into adaptive weighted ranks. Before the transformation, the weight function needs to be optimized based on the distribution of rank changes. For LUSC data, the optimized $\kappa$ was 0.010. Given $ \kappa=0.010$ and $m = 200$, the weights function $-(1-e^\kappa)/(1-e^{-\kappa m}) e^{-\kappa x}$ could be rewritten as $-(1-e^{0.010})/(1-e^{-0.010 \times 200}) e^{-0.010 x}$, with which the original ranks in Table \ref{tab:2} were converted to the weighted ranks in Table \ref{tab:3}. The first derivative of the weights function was negative, indicating a descending trend of the weights, i.e., large weight to the top. The second derivative of the weights function was positive, suggesting more changes in top-ranked genes. For instance, the weighted rank distance between the first and second ranks was $0.0002 \times (0.0115 - 0.0113)$, whereas the value was $0.0001 \times (0.0113 - 0.0112)$ between the second and third genes (Table \ref{tab:3}).  

\begin{table}[]
\centering
\caption{transform the original ranks in table 2 to the weighted ranks using an optimized $\kappa$ of 0.010.}
\label{tab:3}
\resizebox{\columnwidth}{!}{%
\begin{tabular}{llllllllllll}
\toprule
Genes   & Original & Obs 1  & Obs 2  & Obs 3  & Obs 4  & Obs 5  & Obs 6  & Obs 7  & Obs 8  & Obs 9  & Obs 10 \\
\midrule
CDCA8   & 0.0115   & 0.0115 & 0.0115 & 0.0115 & 0.0115 & 0.0115 & 0.0115 & 0.0115 & 0.0115 & 0.0115 & 0.0115 \\
AUNIP   & 0.0113   & 0.0113 & 0.0113 & 0.0113 & 0.0113 & 0.0113 & 0.0113 & 0.0113 & 0.0113 & 0.0113 & 0.0113 \\
KIF2C   & 0.0112   & 0.0112 & 0.0112 & 0.0112 & 0.0112 & 0.0112 & 0.0112 & 0.0112 & 0.0112 & 0.0112 & 0.0112 \\
CDCA5   & 0.0111   & 0.0111 & 0.0110 & 0.0111 & 0.0111 & 0.0111 & 0.0111 & 0.0111 & 0.0111 & 0.0111 & 0.0111 \\
CENPA   & 0.0110   & 0.0110 & 0.0111 & 0.0110 & 0.0110 & 0.0110 & 0.0110 & 0.0110 & 0.0110 & 0.0110 & 0.0110 \\
TPX2    & 0.0109   & 0.0109 & 0.0109 & 0.0109 & 0.0109 & 0.0109 & 0.0109 & 0.0109 & 0.0109 & 0.0109 & 0.0109 \\
TROAP   & 0.0108   & 0.0108 & 0.0106 & 0.0108 & 0.0108 & 0.0108 & 0.0108 & 0.0108 & 0.0108 & 0.0108 & 0.0108 \\
TFAP2A  & 0.0107   & 0.0106 & 0.0104 & 0.0106 & 0.0106 & 0.0106 & 0.0107 & 0.0107 & 0.0107 & 0.0107 & 0.0105 \\
PLK1    & 0.0106   & 0.0107 & 0.0103 & 0.0104 & 0.0107 & 0.0107 & 0.0106 & 0.0106 & 0.0106 & 0.0106 & 0.0106 \\
BIRC5   & 0.0105   & 0.0104 & 0.0105 & 0.0101 & 0.0104 & 0.0105 & 0.0103 & 0.0105 & 0.0104 & 0.0105 & 0.0107 \\
NUF2    & 0.0104   & 0.0103 & 0.0107 & 0.0107 & 0.0103 & 0.0104 & 0.0102 & 0.0104 & 0.0105 & 0.0104 & 0.0101 \\
UBE2C   & 0.0103   & 0.0105 & 0.0099 & 0.0103 & 0.0101 & 0.0101 & 0.0101 & 0.0103 & 0.0102 & 0.0100 & 0.0100 \\
CCNB1   & 0.0102   & 0.0101 & 0.0108 & 0.0102 & 0.0099 & 0.0100 & 0.0100 & 0.0102 & 0.0103 & 0.0102 & 0.0104 \\
RAMP2   & 0.0101   & 0.0100 & 0.0094 & 0.0098 & 0.0102 & 0.0102 & 0.0105 & 0.0100 & 0.0099 & 0.0103 & 0.0102 \\
TTK     & 0.0100   & 0.0099 & 0.0100 & 0.0105 & 0.0100 & 0.0103 & 0.0104 & 0.0101 & 0.0100 & 0.0099 & 0.0099 \\
NOSTRIN & 0.0099   & 0.0098 & 0.0096 & 0.0100 & 0.0105 & 0.0099 & 0.0099 & 0.0099 & 0.0098 & 0.0101 & 0.0098 \\
KIF4A   & 0.0098   & 0.0097 & 0.0098 & 0.0096 & 0.0098 & 0.0097 & 0.0098 & 0.0098 & 0.0101 & 0.0098 & 0.0103 \\
HJURP   & 0.0097   & 0.0096 & 0.0095 & 0.0097 & 0.0096 & 0.0095 & 0.0097 & 0.0097 & 0.0097 & 0.0097 & 0.0097 \\
ACVRL1  & 0.0096   & 0.0095 & 0.0097 & 0.0099 & 0.0097 & 0.0096 & 0.0096 & 0.0096 & 0.0095 & 0.0095 & 0.0096 \\
PRC1    & 0.0095   & 0.0102 & 0.0102 & 0.0094 & 0.0095 & 0.0098 & 0.0095 & 0.0095 & 0.0096 & 0.0096 & 0.0095 \\
\vdots       & \vdots         & \vdots      & \vdots     & \vdots     & \vdots     & \vdots      & \vdots     & \vdots     & \vdots    &\vdots    & \vdots  \\
\bottomrule
\end{tabular}%
}
\end{table}

\subsubsection{Total rank changes (step3)}
\label{3.1.3}

In this step, we computed the overall weighted rank changes for each observation. The total rank changes are the squared sum of weighted rank changes between the original and leave-one-out ranking.

\begin{figure}
    \centering
    \includegraphics[width=1\textwidth]{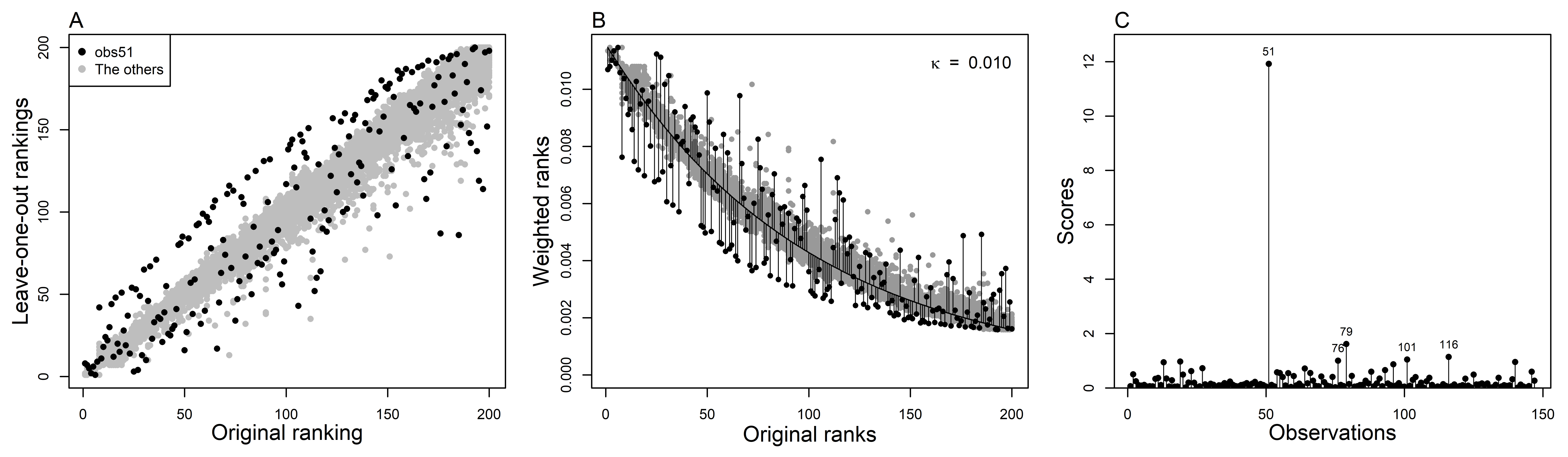}
    \caption{LUSC data. The process of IPs detection. Panel A, derive the original and leave-one-out rankings (step1). The scatter plot depicts the leave-one-out ranks and their corresponding original ranks. Black dots are the leave-one-out ranks of obs51, a potential IP in this data. Panel B, optimize the weight function and transform the original ranks to weighted ranks (step2). The weighted original ranks lie on the line, while their corresponding leave-one-out rank fall above or below the line if rank changes. The vertical distance between the points to the curve is the weighted rank changes (black curve). Panel C, the overall weighted rank changes (step3).}
    \label{fig:2}
\end{figure}

The whole process of IPs detection is visualized in Figure \ref{fig:2}. Panel A exhibits the distribution of rank changes (points not on the diagonal) in the original scale, where more rank changes are observed on the tail side than on the head (step 1). In the second step, original ranks were transformed to weighted ranks with a weight function optimized based on the distribution of the rank changes (Panel B). The weight function curve will be steeper if more changes are in the tail, where is allocated with less weights. In contrast, the weight function will allocate more weights to the tail, if rank changes are more evenly distributed between the head and tail.

In Panel B, points without rank changes will lie on the curve, while the points that fall above or below are the ones with observed rank changes. The vertical lines connecting points to the line represent the weighted rank changes in the absolute form when obs51 is removed.

In the third step, we calculated the total rank changes of each observation (Panel Panel C exhibited the total rank changes for each observation. Obs51 is identified as a potential IP, considering the large gap with the rest cases. In fact, the abnormal effects of obs51 can also be observed in panel A and panel B (see the black points). This example highlights that a single case can severely influence the feature rankings. 

\subsection{The choice of weights}
\label{sec:3.2}

\begin{figure}
    \centering
    \includegraphics[width=1\textwidth]{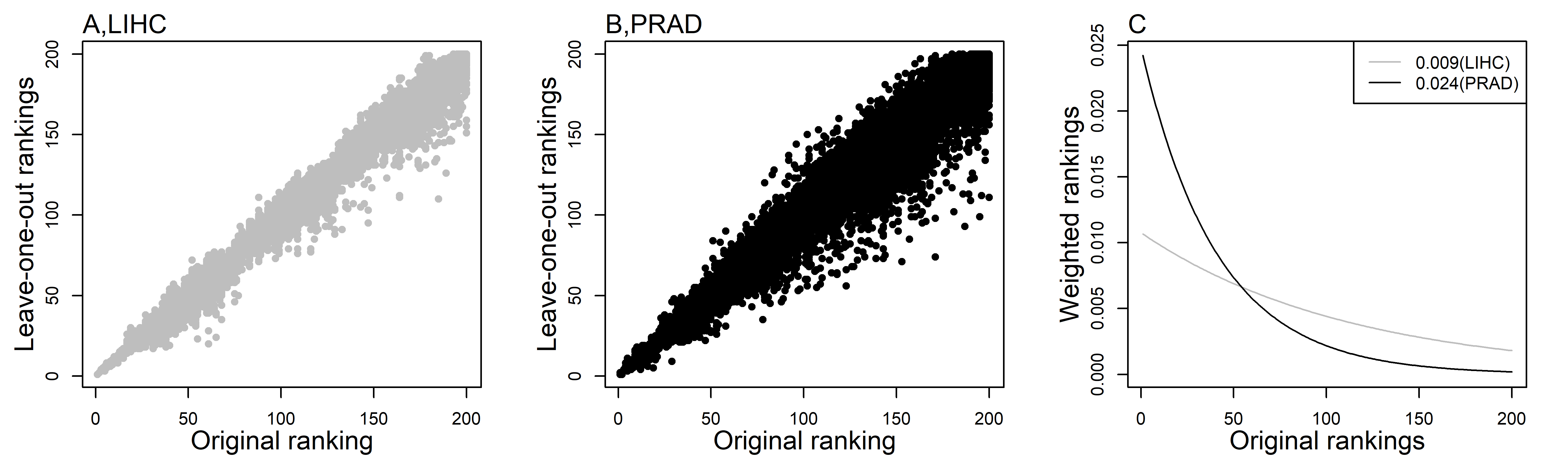}
    \caption{the adaptive weights for LIHC and PRAD data. A, rank changes are more evenly distributed; B, more rank changes in the tail side. C, the optimized weights for the two data. }
    \label{fig:3}
\end{figure}

The choice of weights depends on the distribution of rank changes. We illustrated this by comparing the feature rankings of two datasets, LIHC and PRAD. The rank changes for LIHC are more evenly distributed; therefore, the optimized weight function assigns fewer weights to the top, inducing the reduced total rank changes (Figure \ref{fig:3}A and \ref{fig:3}C. In contrast, PRAD observes more rank changes in the tail, therefore, allocating fewer weights to the tail (Figure \ref{fig:3}B and \ref{fig:3}C). Figure \ref{fig:3}C shows that the optimized $\kappa$ is 0.009 for LIHC and 0.024 for PRAD, and a larger value indicates more weights to the top. The comparison of the two datasets intuitively showed that the adaptive weight method is adjusted to better fit the distribution of rank changes; Therefore, the method is more flexible and can be tailored to the data being analyzed. 

\subsection{Influential points}
\label{sec:3.3}
This study utilized graphical displays to identify potential IPs. Hawkins stated that “a sample containing outliers would show such characteristics as significant gaps between ‘outlying’ and ‘inlying’ observations.” As such, a broad gap is an indicator of potential IPs \cite{Hawkins.1980}. For instance, Obs51 is a poential IP in LUSC, considering the evident gap between obs51 and the rest. Similar gaps can also be found in obs22 in STAD (Figure \ref{fig:4}), obs73 in COAD, obs93 in HNSC, obs107 in PRAD, and obs134 in THCA (Figure A1 in Appendix). Such gaps cannot be observed in LIHC and LUAD, indicating no outstanding IP. Interpretation of the results can be complex when multiple extreme cases have similar weighted rank changes. In BRCA, both obs53 and obs338 are likely to be the potential IPs. As aforementioned, we can detect at most one potential IP due to the restriction of leave-one-out. In this case, a simple interpretation is that obs338 is still a candidate IP; however, multiple IPs may exist. The check for multiple IPs is a complex topic that has been beyond the scope of this research.

It is worth noting that strategies for detecting influential points can only offer indications of potential influential points, rather than conclusive evidence for definitive diagnosis.  A comprehensive investigation into the causes of the anomaly is necessary before a final judgment can be made \cite{Aguinis.2013}.

\begin{figure}
    \centering
    \includegraphics[width=1\textwidth]{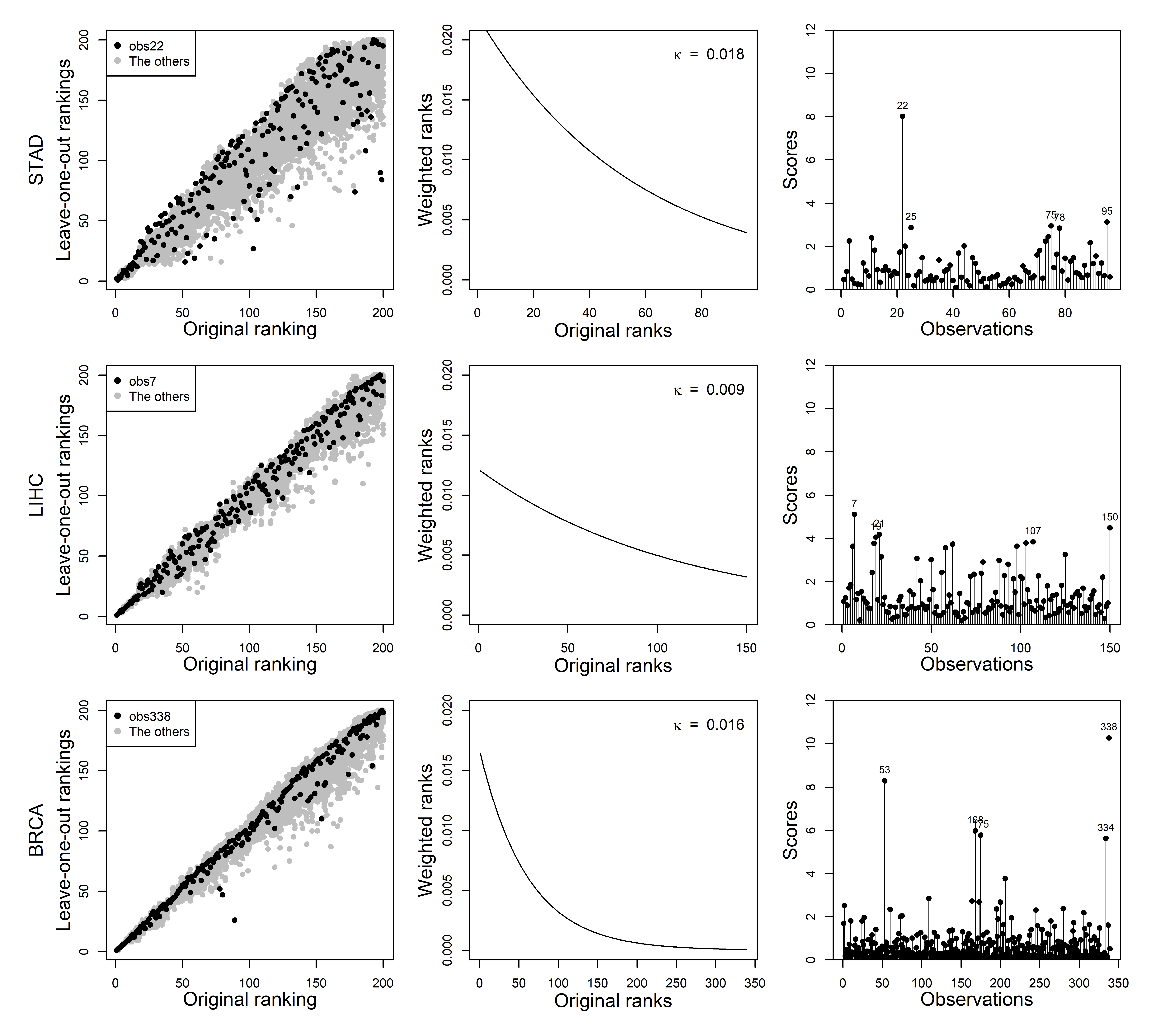}
    \caption{the IPs detection procedure for STAD, LIHC, and BRCA data.  
From left to right are original rankings, weighted ranking, and total weighted rank changes. }
    \label{fig:4}
\end{figure}

\subsection{Compare with the other methods}
\label{sec:3.4}

We compared our proposal to the other two methods: the usual Spearman’s coefficient and the weighted extension. In most cases, the three methods are consistent in identifying the outstanding IPs, such as obs22 in STAD (Figure \ref{fig:5}), obs73 in COAD, obs51 in LUSC, and so on (Figure A2 in appendix). There are also cases where the three methods identified different IPs, such as BRCA and LUAD data. We further explored the reasons for the difference between the three methods in BRCA data, where obs175, obs53, and obs338 are the most influential cases using the unweighted, weighted, and our adaptive method, respectively (Figure \ref{fig:5}). 

\begin{figure}
    \centering
    \includegraphics[width=1\textwidth]{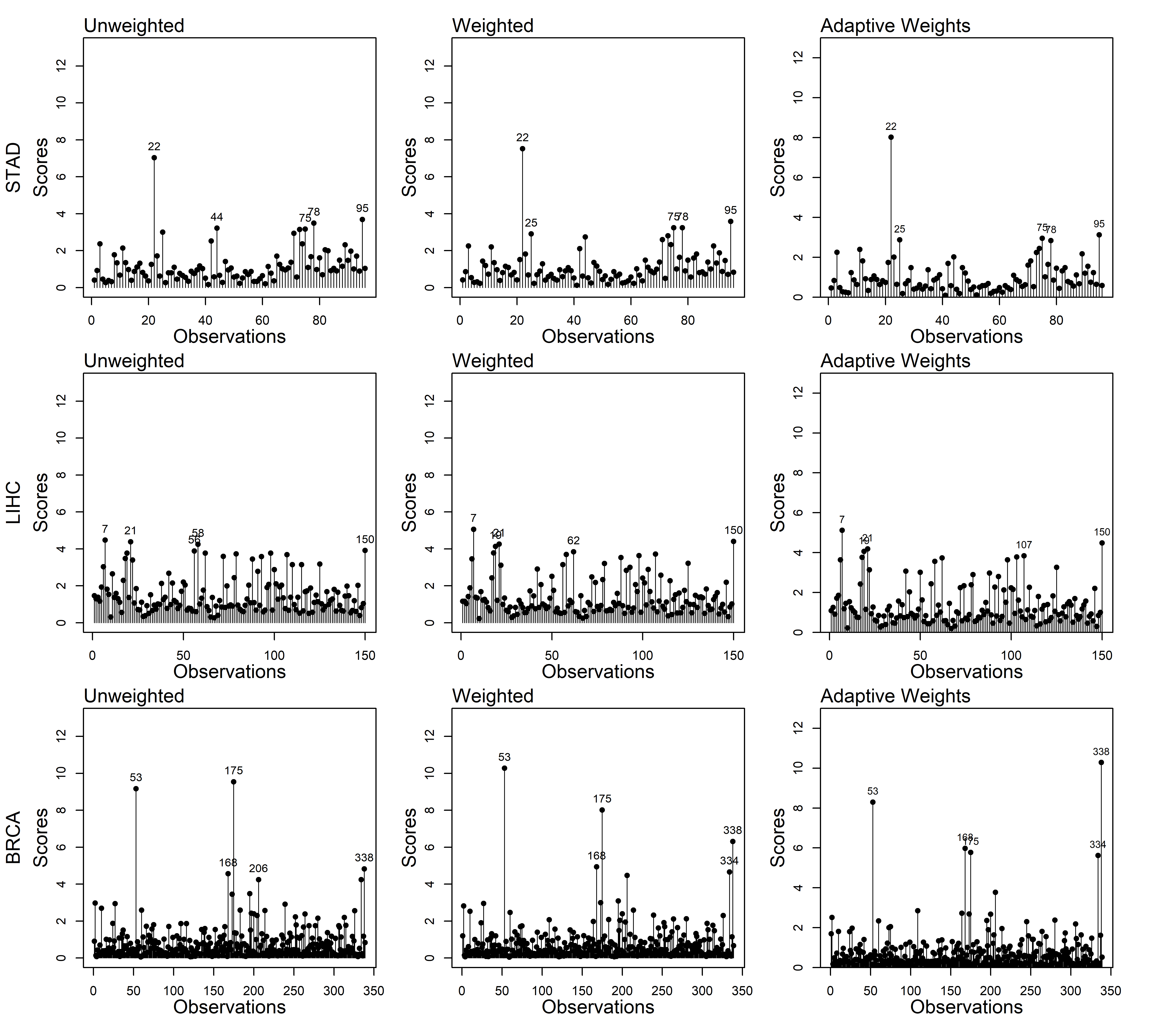}
    \caption{STAD, LIHC, and BRCA data. Compare the results of the three rank comparison methods. From left to right are the results of Spearman’s coefficient, weighted Spearman’s coefficient, and rank comparison using adaptive weights. }
    \label{fig:5}
\end{figure}

\begin{figure}
    \centering
    \includegraphics[width=1\textwidth]{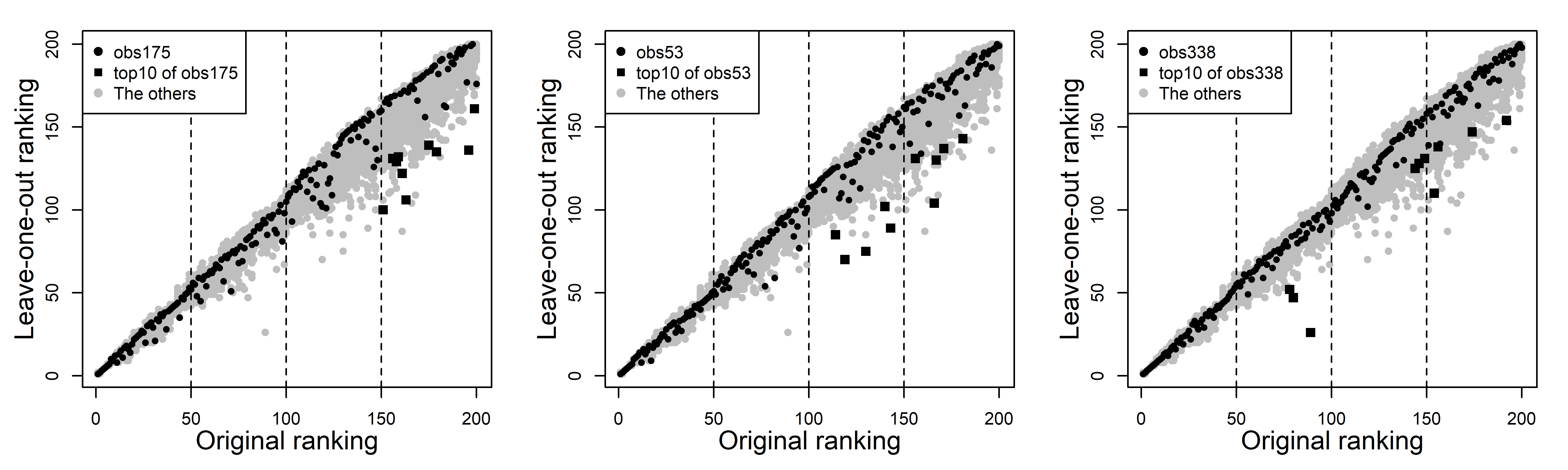}
    \caption{BRCA data. The rank changes caused by the three potential IPs (obs175, obs53, and obs338) identified with three rank comparison methods. The top10 changed features are marked with squares. From left to right to right are the IPs detected with Spearman’s coefficient, weighted Spearman’s coefficient, and rank comparison using adaptive weights. }
    \label{fig:6}
\end{figure}

Figure \ref{fig:6} depicts the influence of the three potential IPs. For obs175, the top 10 changed features are found in the tail side (rank 150-200 in the original ranking). The deletion of obs53 results in more changes between 100 and 150 (in the original ranking) since it weights more toward the top features. The adaptive weights identify obs338 as the most influential case, whose deletion may not produce the largest overall rank changes but identifies the changes of three important genes ranked between 50 and 100. The adaptive weights allocate more weights to the top in this data, which we believe is more appropriate compared with the weighted Spearman's coefficient that uses fixed weights.

\subsection{The effects of IPs on subsequent analyses: example with GSEA}
\label{sec:3.5}

\begin{figure}
    \centering
    \includegraphics[width=0.9\textwidth]{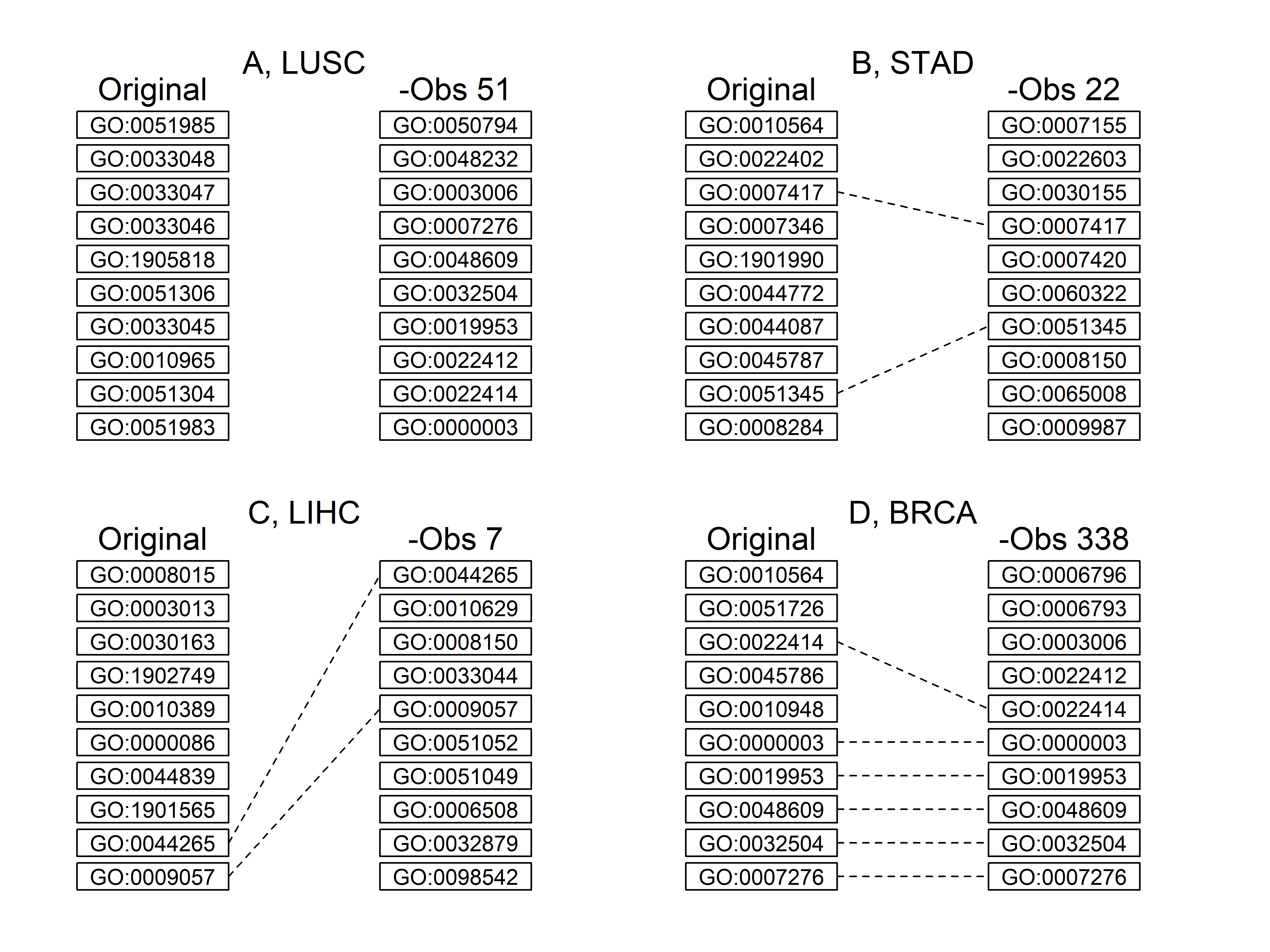}
    \caption{the effects of influential points on subsequent analyses, using GSEA as an example.For each data, the two columns are the top10 pathways derived from GSEA using the original data and data with a potential IP removed. The pathways presented in both were linked with dashed lines. }
    \label{fig:7}
\end{figure}

In the above, we have demonstrated the impacts of IPs on feature rankings, but have not yet shown the potential impacts of using a distorted feature ranking on subsequent analyses. Here, we investigated such effects on GSEA, a routine downstream analysis that utilizes feature ranking as input. Obs51 is an IP in LUSC, and its removal results in substantial alteration in the top 10 pathways, indicating a severe influence on the results (Figure \ref{fig:7}). Similar results can also be found in THCA, where obs134 is a potential IP (Figure A3). The removal of Obs22 in STAD, and Obs7 in LIHC only yields one and two shared pathways. BRCA, LUAD, and KIRC seem to be less influenced by IPs, out of which LUAD and KIRC do not contain IP, while BRCA identifies obs338 as a potential IP. Briefly, the presence of IPs can severely affect the subsequent analyses.

\section{Discussion}

This study presents a novel approach to identify influential points (IPs) in feature rankings using leave-one-out and a rank comparison method with adaptive weights. With the methods, we detected potential IPs in several widely-used datasets, indicating the necessity of IPs detection when using these data. A single IP can severely distort the feature rankings, ultimately affects the subsequent analyses. LUSC data was a representative example showing the severe effects of a single case. However, previous researches using this data did not detect, handle, or report the IP in their analyses \cite{Sun.2017}. we hope that our study will raise awareness about the importance of identifying IPs when deriving feature rankings.

This study proposed a weighted rank comparison method used in IPs detection. The rank comparison method uses adaptive weights based on the distribution of rank changes. This can avoid the potential bias caused by using arbitrary weights. Compared with other methods, the adaptive weights exhibited favorable characteristics in identifying the rank changes of important features. 

The effects of IPs on feature rankings are heavily underrated. As far as I know, this study probably is the first work concentrating on IPs detection on feature rankings. We claimed more attention to this issue to enhence the quality of researches \cite{Boulesteix.2009}. 

\section*{Acknowledgments}
The results shown here were based on  data provided by the TCGA Research Network: \url{https://www.cancer.gov/tcga}. 

\bibliographystyle{unsrt}  
\bibliography{references}  

\begin{thebibliography}{10}

\bibitem{Churko.2013}
Jared~M. Churko, Gary~L. Mantalas, Michael~P. Snyder, and Joseph~C. Wu.
\newblock Overview of high throughput sequencing technologies to elucidate
  molecular pathways in cardiovascular diseases.
\newblock {\em Circulation research}, 112(12):1613--1623, 2013.

\bibitem{WietlickaPiszcz.2013}
Magdalena Wietlicka-Piszcz.
\newblock The stability of gene selection in microarray experiments.
\newblock {\em Studies in Logic, Grammar and Rhetoric}, 35(1):87--101, 2013.

\bibitem{Boulesteix.2009}
Anne-Laure Boulesteix and Martin Slawski.
\newblock Stability and aggregation of ranked gene lists.
\newblock {\em Briefings in bioinformatics}, 10(5):556--568, 2009.

\bibitem{Li.2011}
Qunhua Li, James~B. Brown, Haiyan Huang, and Peter~J. Bickel.
\newblock Measuring reproducibility of high-throughput experiments.
\newblock {\em The annals of applied statistics}, 5(3):1752--1779, 2011.

\bibitem{Shi.2007}
Jing Shi and Michael~G. Walker.
\newblock Gene set enrichment analysis (gsea) for interpreting gene expression
  profiles.
\newblock {\em Current Bioinformatics}, 2(2):133--137, 2007.

\bibitem{Aerts.2006}
Stein Aerts, Diether Lambrechts, Sunit Maity, Peter {van Loo}, Bert Coessens,
  Frederik de~Smet, Leon-Charles Tranchevent, Bart de~Moor, Peter Marynen, and
  Bassem Hassan.
\newblock Gene prioritization through genomic data fusion.
\newblock {\em Nature biotechnology}, 24(5):537--544, 2006.

\bibitem{Pham.2006}
Tuan~D. Pham, Christine Wells, and Denis~I. Crane.
\newblock Analysis of microarray gene expression data.
\newblock {\em Current Bioinformatics}, 1(1):37--53, 2006.

\bibitem{Heydebreck.2004}
Anja von Heydebreck, Wolfgang Huber, and Robert Gentleman.
\newblock Differential expression with the bioconductor project.
\newblock 2004.

\bibitem{Subramanian.2005}
Aravind Subramanian, Pablo Tamayo, Vamsi~K. Mootha, Sayan Mukherjee,
  Benjamin~L. Ebert, Michael~A. Gillette, Amanda Paulovich, Scott~L. Pomeroy,
  Todd~R. Golub, and Eric~S. Lander.
\newblock Gene set enrichment analysis: a knowledge-based approach for
  interpreting genome-wide expression profiles.
\newblock {\em Proceedings of the National Academy of Sciences},
  102(43):15545--15550, 2005.

\bibitem{Nurunnabi.2014}
A.~A.M. Nurunnabi, Ali~S. Hadi, and AHMR Imon.
\newblock Procedures for the identification of multiple influential
  observations in linear regression.
\newblock {\em Journal of Applied Statistics}, 41(6):1315--1331, 2014.

\bibitem{He.2010}
Zengyou He and Weichuan Yu.
\newblock Stable feature selection for biomarker discovery.
\newblock {\em Computational biology and chemistry}, 34(4):215--225, 2010.

\bibitem{Alelyani.}
Salem Alelyani, Zheng Zhao, and Huan Liu.
\newblock A dilemma in assessing stability of feature selection algorithms.
\newblock In {\em 2011 IEEE International Conference on High Performance
  Computing and Communications}, pages 701--707. IEEE.

\bibitem{Zyla.2017}
Joanna Zyla, Michal Marczyk, January Weiner, and Joanna Polanska.
\newblock Ranking metrics in gene set enrichment analysis: do they matter?
\newblock {\em BMC bioinformatics}, 18(1):1--12, 2017.

\bibitem{Gawe.2013}
Danuta Gawe{\l} and Krzysztof Fujarewicz.
\newblock On the sensitivity of feature ranked lists for large-scale biological
  data.
\newblock {\em Mathematical Biosciences {\&} Engineering}, 10(3):667, 2013.

\bibitem{Li.2019}
Xue Li, Xinlei Wang, and Guanghua Xiao.
\newblock A comparative study of rank aggregation methods for partial and top
  ranked lists in genomic applications.
\newblock {\em Briefings in bioinformatics}, 20(1):178--189, 2019.

\bibitem{Anders.2012}
Simon Anders and Wolfgang Huber.
\newblock Differential expression of rna-seq data at the gene level--the deseq
  package.
\newblock {\em Heidelberg, Germany: European Molecular Biology Laboratory
  (EMBL)}, 10:f1000research, 2012.

\bibitem{Zimek.2012}
Arthur Zimek, Erich Schubert, and Hans-Peter Kriegel.
\newblock A survey on unsupervised outlier detection in high--dimensional
  numerical data.
\newblock {\em Statistical Analysis and Data Mining: The ASA Data Science
  Journal}, 5(5):363--387, 2012.

\bibitem{Everitt.2002}
Brian Everitt and Anders Skrondal.
\newblock {\em The Cambridge dictionary of statistics}, volume 106.
\newblock {Cambridge university press Cambridge}, 2002.

\bibitem{Cook.1977}
R.~Dennis Cook.
\newblock Detection of influential observation in linear regression.
\newblock {\em Technometrics}, 19(1):15--18, 1977.

\bibitem{Belsley.2005}
David~A. Belsley, Edwin Kuh, and Roy~E. Welsch.
\newblock {\em Regression diagnostics: Identifying influential data and sources
  of collinearity}, volume 571.
\newblock {John Wiley {\&} Sons}, 2005.

\bibitem{Hutter.2018}
Carolyn Hutter and Jean~Claude Zenklusen.
\newblock The cancer genome atlas: creating lasting value beyond its data.
\newblock {\em Cell}, 173(2):283--285, 2018.

\bibitem{Chen.2014}
Yunshun Chen, Davis McCarthy, Mark Robinson, and Gordon~K. Smyth.
\newblock edger: differential expression analysis of digital gene expression
  data user's guide.
\newblock {\em Bioconductor User's Guide}, 2014.

\bibitem{Cui.2003}
Xiangqin Cui and Gary~A. Churchill.
\newblock Statistical tests for differential expression in cdna microarray
  experiments.
\newblock {\em Genome biology}, 4(4):1--10, 2003.

\bibitem{Robinson.2010}
Mark~D. Robinson, Davis~J. McCarthy, and Gordon~K. Smyth.
\newblock edger: a bioconductor package for differential expression analysis of
  digital gene expression data.
\newblock {\em Bioinformatics}, 26(1):139--140, 2010.

\bibitem{Anders.2010}
Simon Anders and Wolfgang Huber.
\newblock Differential expression analysis for sequence count data.
\newblock {\em Nature Precedings}, page~1, 2010.

\bibitem{Law.2014}
Charity~W. Law, Yunshun Chen, Wei Shi, and Gordon~K. Smyth.
\newblock voom: Precision weights unlock linear model analysis tools for
  rna-seq read counts.
\newblock {\em Genome biology}, 15(2):1--17, 2014.

\bibitem{Leng.2013}
Ning Leng, John~A. Dawson, James~A. Thomson, Victor Ruotti, Anna~I. Rissman,
  Bart M.~G. Smits, Jill~D. Haag, Michael~N. Gould, Ron~M. Stewart, and
  Christina Kendziorski.
\newblock Ebseq: an empirical bayes hierarchical model for inference in rna-seq
  experiments.
\newblock {\em Bioinformatics}, 29(8):1035--1043, 2013.

\bibitem{Rajaratnam.2019}
Bala Rajaratnam, Steven Roberts, Doug Sparks, and Honglin Yu.
\newblock Influence diagnostics for high-dimensional lasso regression.
\newblock {\em Journal of Computational and Graphical Statistics},
  28(4):877--890, 2019.

\bibitem{Abdi.2007}
Herv{\'e} Abdi.
\newblock The kendall rank correlation coefficient.
\newblock {\em Encyclopedia of Measurement and Statistics. Sage, Thousand Oaks,
  CA}, pages 508--510, 2007.

\bibitem{Sedgwick.2014}
Philip Sedgwick.
\newblock Spearman's rank correlation coefficient.
\newblock {\em Bmj}, 349, 2014.

\bibitem{Stillwell.1981}
William~G. Stillwell, David~A. Seaver, and Ward Edwards.
\newblock A comparison of weight approximation techniques in multiattribute
  utility decision making.
\newblock {\em Organizational behavior and human performance}, 28(1):62--77,
  1981.

\bibitem{Solymosi.1986}
Tam{\'a}s Solymosi and Jozsef Dombi.
\newblock A method for determining the weights of criteria: the centralized
  weights.
\newblock {\em European journal of operational research}, 26(1):35--41, 1986.

\bibitem{Yang.2006}
Xinan Yang, Stefan Bentink, Stefanie Scheid, and Rainer Spang.
\newblock Similarities of ordered gene lists.
\newblock {\em Journal of bioinformatics and computational biology},
  4(03):693--708, 2006.

\bibitem{DaPintoCosta.2005}
Joaquim {Da Pinto Costa} and Carlos Soares.
\newblock A weighted rank measure of correlation.
\newblock {\em Australian {\&} New Zealand Journal of Statistics},
  47(4):515--529, 2005.

\bibitem{Hawkins.1980}
Douglas~M. Hawkins.
\newblock {\em Identification of outliers}, volume~11.
\newblock Springer, 1980.

\bibitem{Aguinis.2013}
Herman Aguinis, Ryan~K. Gottfredson, and Harry Joo.
\newblock Best-practice recommendations for defining, identifying, and handling
  outliers.
\newblock {\em Organizational Research Methods}, 16(2):270--301, 2013.

\bibitem{Sun.2017}
Fenghao Sun, Xiaodong Yang, Yulin Jin, Li~Chen, Lin Wang, Mengkun Shi, Cheng
  Zhan, Yu~Shi, and Qun Wang.
\newblock Bioinformatics analyses of the differences between lung
  adenocarcinoma and squamous cell carcinoma using the cancer genome atlas
  expression data.
\newblock {\em Molecular medicine reports}, 16(1):609--616, 2017.

\end{thebibliography}

\pagebreak

\appendix
\counterwithin{figure}{section}
\section{Appendix Figures}

\begin{figure}[htp]
    \centering
    \includegraphics[width = 0.65\textwidth]{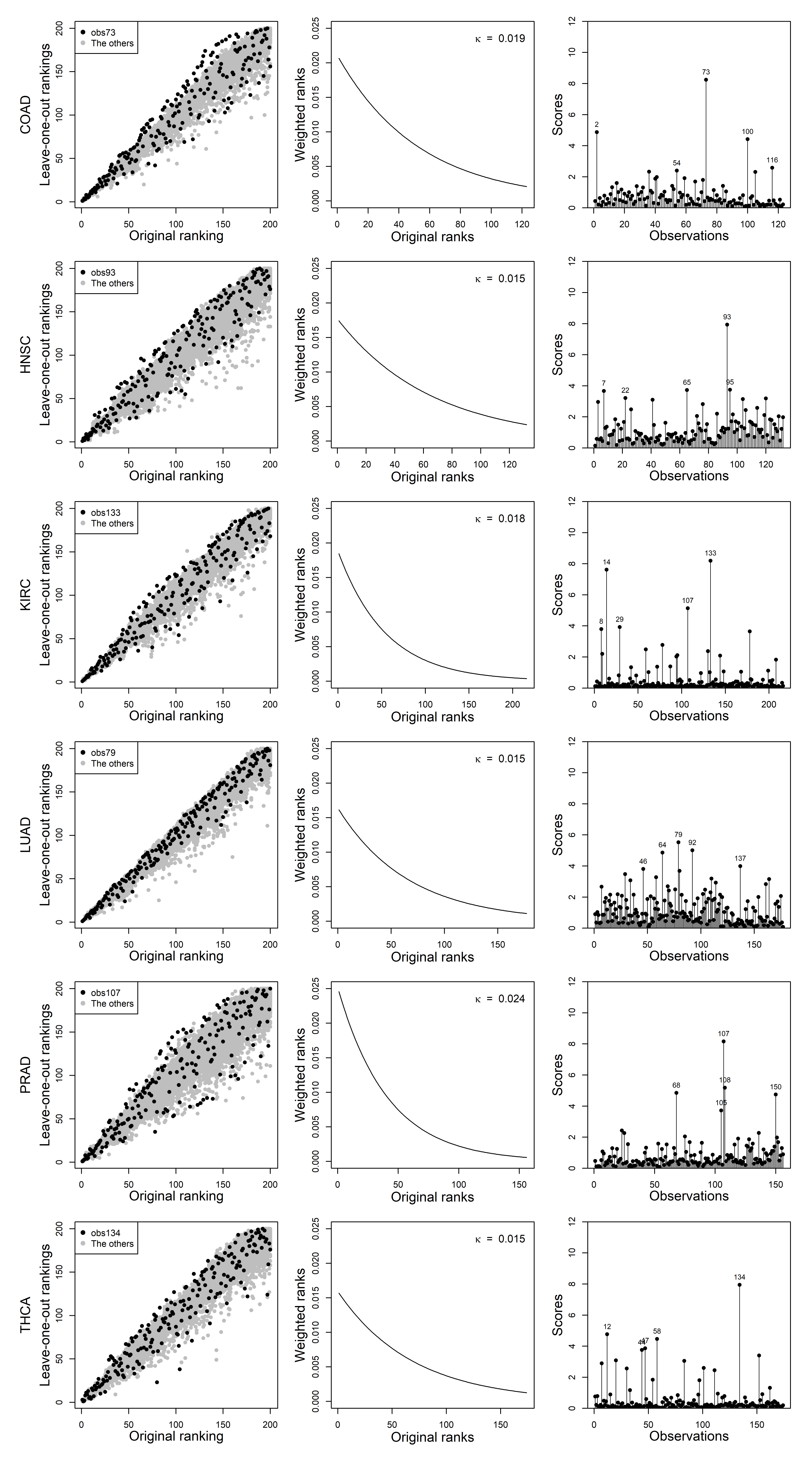}
    \caption{The potential IPs and the adaptive weight function}
    \label{fig:A1}
\end{figure}

\begin{figure}[htp]
    \centering
    \includegraphics[width=0.7\textwidth]{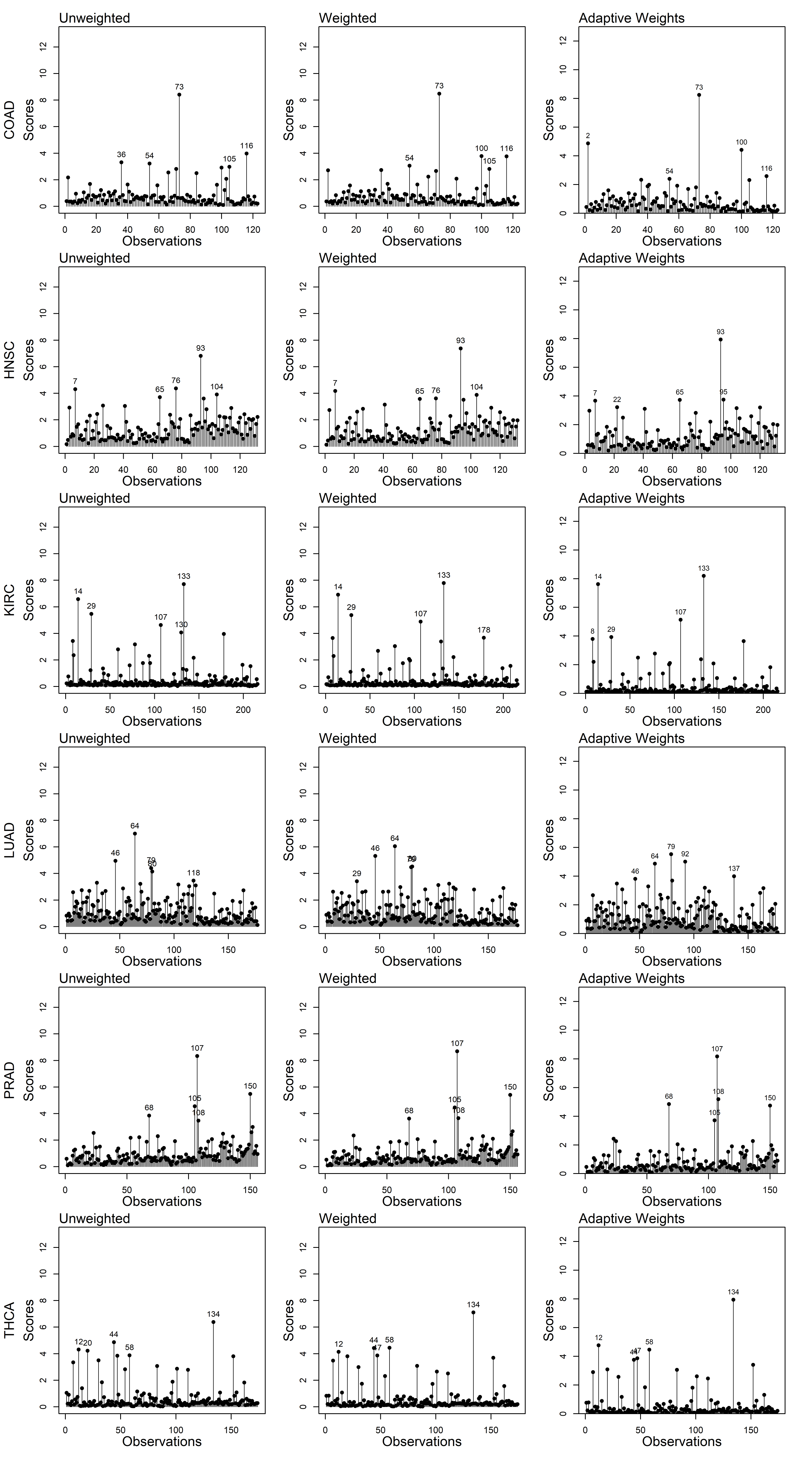}
    \caption{Compare the results of the three rank comparison methods. From the left to the right are the results of the Spearman’s coefficient, weighted Spearman’s coefficient, and rank comparison using adaptive weights.}
    \label{fig:A2}
\end{figure}

\begin{figure}[htp]
    \centering
    \includegraphics[width=0.8\textwidth]{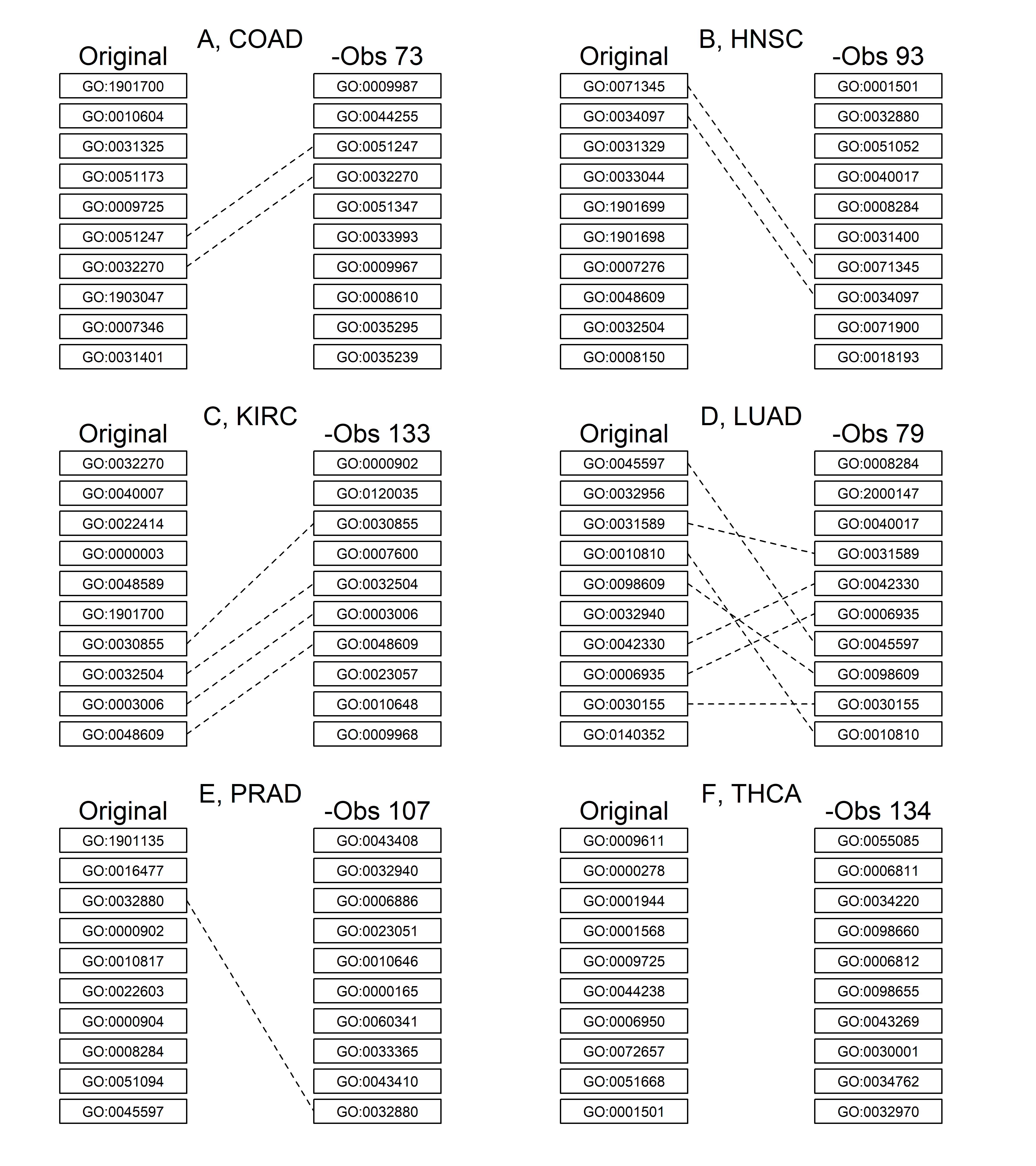}
    \caption{The effects of IPs on GSEA.}
    \label{fig:A3}
\end{figure}

\end{document}